\documentclass[11pt]{article}
\usepackage{moriond}
\usepackage{xspace}
\usepackage{subfigure}

\def\Journal#1#2#3#4{{#1} {\bf #2}, #3 (#4)}

\def\PLB{{\em Phys. Lett.}  B}
\def\PRL{\em Phys. Rev. Lett.}
\def\PRD{{\em Phys. Rev.} D}

\def\JHEP{JHEP}
\def\EPJ{EPJ}

\def\be{\begin{equation}}
\def\ee{\end{equation}}
\def\bea{\begin{eqnarray}}
\def\eea{\end{eqnarray}}

\def\ttbar{\ensuremath{\bar{t}t}\xspace}
\def\ff{\ensuremath{4^{{th}}}\xspace}
\def\de{\ensuremath{\delta}\xspace}

\newcommand{\eslash}{\ensuremath{{\hbox{$E$\kern-0.6em\lower-.05ex\hbox{/}\kern0.10em}}}}

\def\met{\ensuremath{\eslash_{\rm T}}\xspace}
\def\pp{\ensuremath{\mbox{pp}}\xspace}
\def\fbinv{\ensuremath{\mbox{fb}^{-1}}\xspace}
\def\pt{\ensuremath{p_{T}}\xspace}

\def\MD{\ensuremath{M_{{D}}}\xspace}
\def\ee{\ensuremath{\mbox{ee}}\xspace}
\def\mm{\ensuremath{\mu\mu}\xspace}
\def\MS{\ensuremath{M_{{S}}}\xspace}

\def\seighttev{\ensuremath{\sqrt{s}=8} TeV\xspace}
\def\sseventev{\ensuremath{\sqrt{s}=7} TeV\xspace}
\def\HT{\ensuremath{H_{{T}}}\xspace}
\def\ST{\ensuremath{S_{{T}}}\xspace}

\newcommand{\Photo}{}

\begin{document}
\vspace*{4cm}
\title{Search for extra-dimensions, \ttbar resonances, \ff generation and leptoquark signatures at the LHC}

\author{ Matthieu Marionneau \\
(On behalf of the CMS and ATLAS collaborations) }

\address{University of Maryland,\\
College Park, USA}

\maketitle

\begin{abstract}
The searches for extra-dimensions, top resonances, \ff generation quarks and leptoquark signatures are presented. The results are based on proton-proton collision data at $\sqrt{s}=7$ or 8 TeV, corresponding to various integrated luminosities. No signal of physics beyond the Standard Model has been observed so far.
\end{abstract}

\section{Introduction}

In this paper are presented a selection of searches of new physics beyond the Standard Model performed by the ATLAS \cite{ATLAS} and CMS \cite{CMS} collaborations, with proton-proton collision data recorded in 2011 and 2012 at \sseventev and \seighttev. Physics model covered by this paper are various, from extra-dimensions searches to black holes, dark matter, high mass resonances involving top quarks, \ff generation quark and leptoquarks searches. SuperSymmetry and other exotic physics searches are discussed in other proceedings of the Moriond EWK 2013 conference.

\section{Large extra-dimensions searches}

The search for large compactified extra-dimensions (EDs), as proposed by the Arkani-Hamed, Dimopoulos and Dvali (ADD) model \cite{ADD}, is motivated by the solve of the hierarchy problem of the SM referring to the large difference between the electroweak scale ($M_{{EWK}}\sim 1$ TeV) and the gravity scale ($M_{planck}\sim 10^{15}$ TeV). In the ADD model, the gravity propagates into a 3+1+\de dimension space, where $\de \geq 2 $ is the number of EDs, while the Standard Model particles are confined into a 3+1 dimension space. In this framework the gravitational flux is diluted in this multidimensional space and the effective Planck scale $M_{{D}}$ can be reduced to a scale similar to $M_{{EWK}}$ and gravitons can be produced at the LHC. Several experimental signatures indicating the presence of EDs have been studied by the ATLAS and CMS collaboration.



\subsection{Graviton production searches}
\label{sec:EDGrav}
\begin{figure}
  \begin{center}
    \subfigure[]{\includegraphics[width=0.55\linewidth]{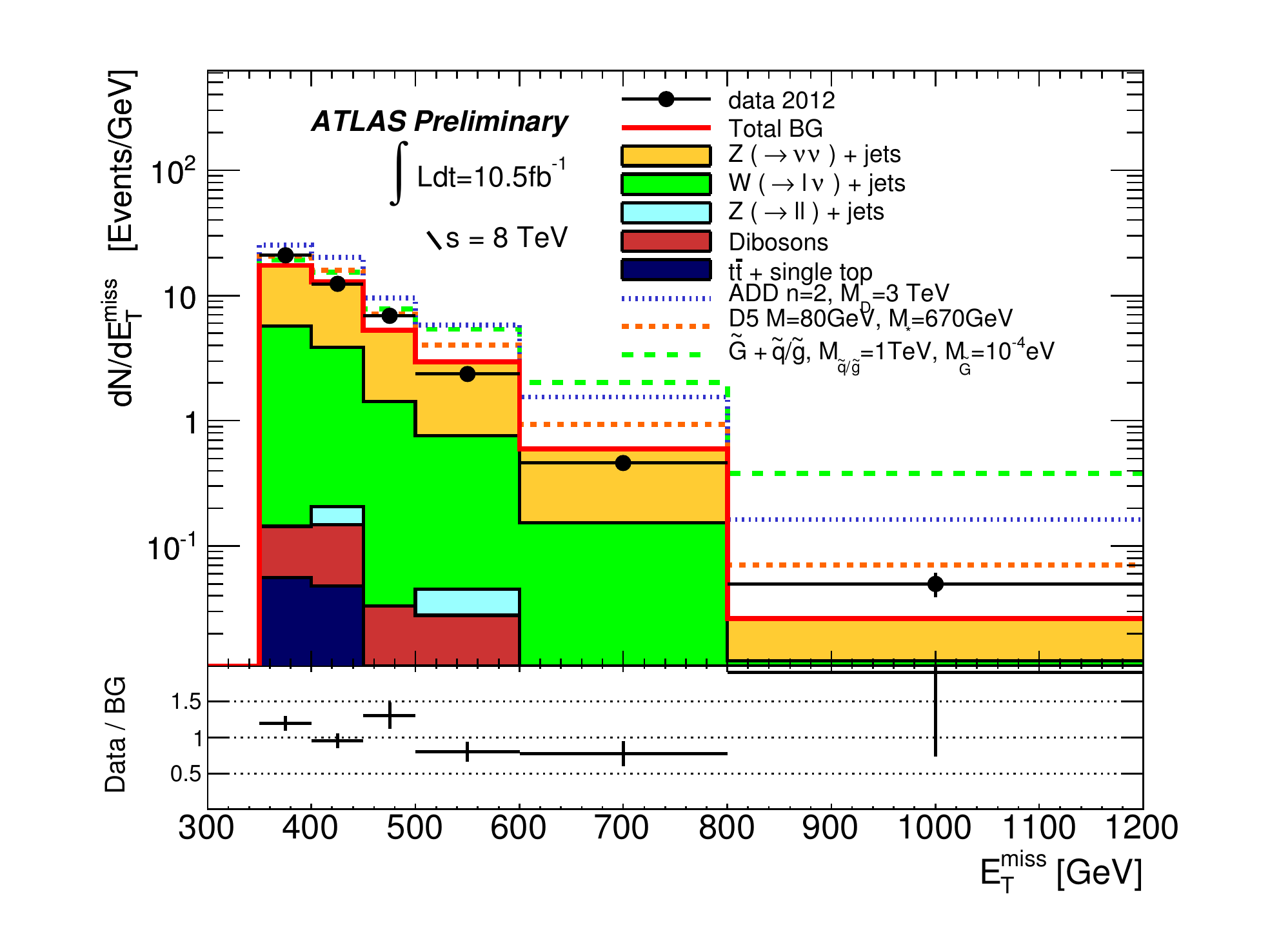}}
    \subfigure[]{\includegraphics[width=0.40\linewidth]{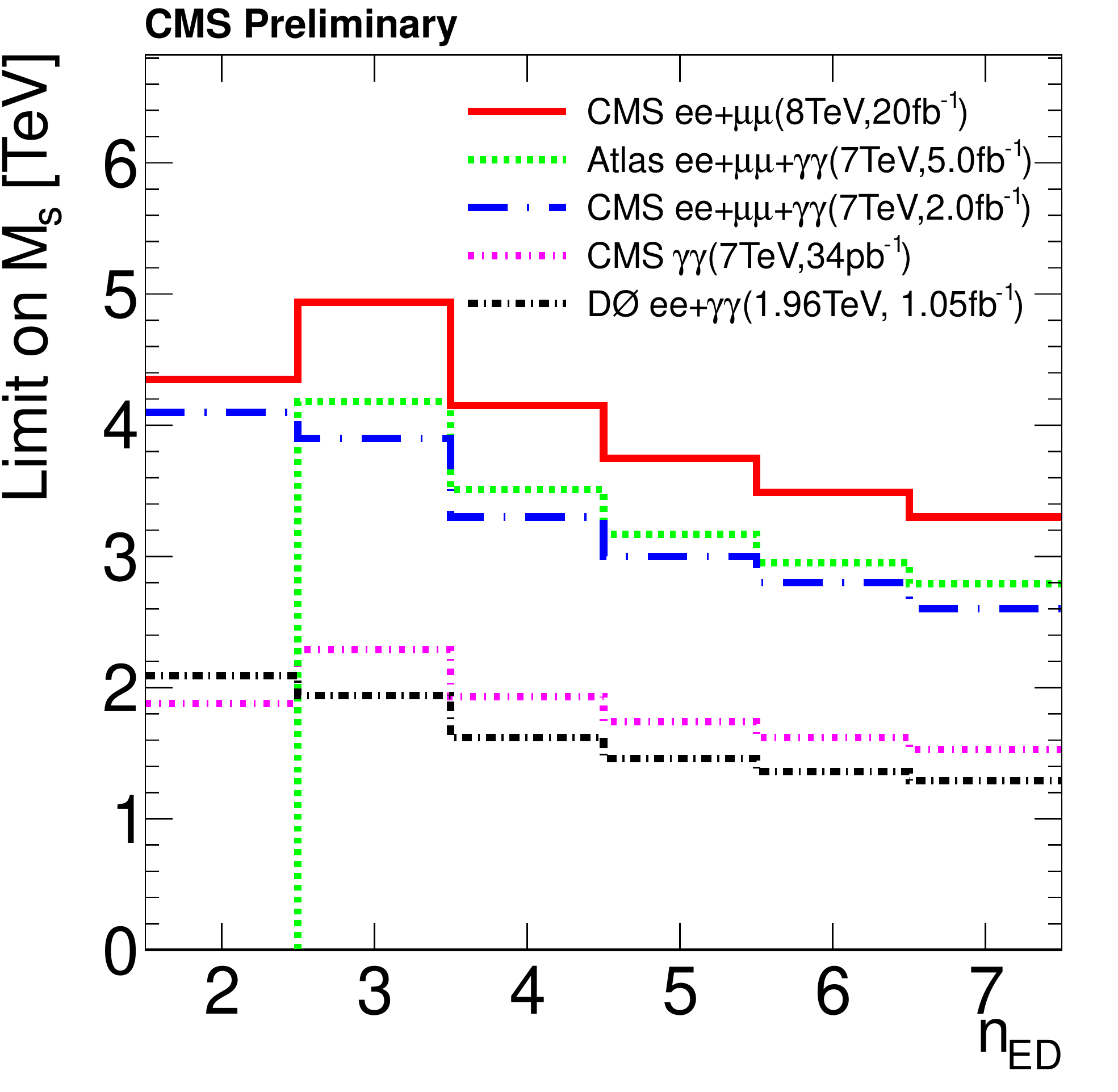}}
    \caption{(a) \met spectrum of events selected in a jet+\met final state by the ATLAS experiment (b) summary of limits on \MS as a function of \de, for several experiments and data samples.}
    \label{fig:ExtraDim}
  \end{center}
\end{figure}

Since the graviton is able to propagate in the extra-dimensions, the first type of signature indicating the presence of EDs is a large amount of missing transverse energy (\met). The final state jet+\met has been performed by the ATLAS experiment with a sample of \pp data at \seighttev corresponding to 10 \fbinv \cite{MJATLAS} \footnote{Only the most recent analyses are discussed here. Several analyses are then intentionally not discussed in this proceeding.}. The offline event selection requires a large amount of \met, one high-\pt jet and no additional lepton or jet in the event. Several filters are applied to reject beam-related backgrounds, cosmic rays and anomalous electronic-noise events. 
 Fig. \ref{fig:ExtraDim}.a shows the \met spectrum for the selected events. The number of observed events in the data is in good agreement with the SM expectations and lower limits are set on the ADD model variable \MD, as a function of \de: values of \MD are excluded below 2.58 ($\de=2$) to 2.88 ($\de=6$) TeV at 95\% C.L.. \\

Other studies have been performed by looking for an excess of data at high dilepton masses, indicating the presence of gravitons decaying into a lepton-antilepton pair. The CMS collaboration studied both \ee and \mm decay channels with \seighttev \pp data corresponding to a total integrated luminosity of 20 \fbinv \cite{DIECMS,DIMCMS}. Events are selected by requiring the presence of two well-identified, opposite charge, same flavor, isolated leptons. Since no excess at high masses have been noticed in both channels, limits are set on \MS, related to \MD by the formula $\MS = 2\sqrt{\pi} \left[ \Gamma(\delta /2 ) \right]^{1/(\delta+2)} \MD$. Fig. \ref{fig:ExtraDim}.b illustrates a summary of the lower limits set on \MS by several experiments for $\de \in [2,7]$.

\subsection{Black holes}

\begin{figure}
  \begin{center}
    \subfigure[]{\includegraphics[width=0.45\linewidth]{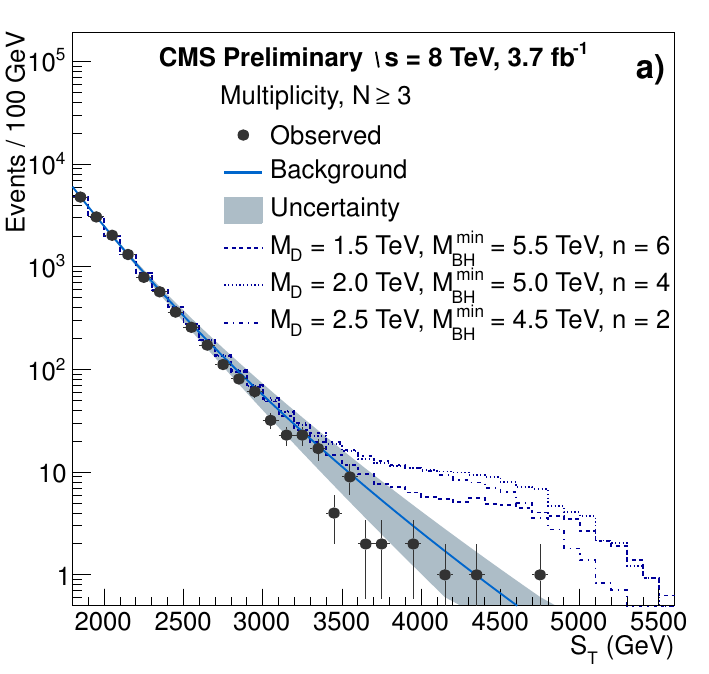}}
    \subfigure[]{\includegraphics[width=0.45\linewidth]{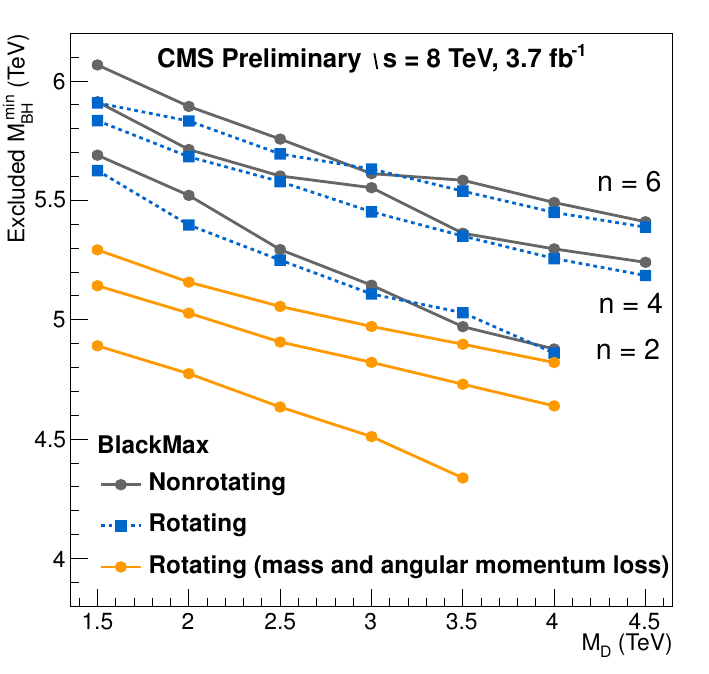}}
    \caption{(a) \ST spectrum for an object multiplicity $N=3$. Black points correspond to data, blue line to SM backgrounds and dotted liens to black holes signal for several benchmarks (b) lower limits on black hole masses for different theoretical models. }
    \label{fig:BH}
  \end{center}
\end{figure}

A consequence of the presence of extra-dimensions is the production of microscopic black-holes in \pp collisions at the LHC. The experimental signature of a microscopic black hole is a large transverse energy carried by physical objects (charged leptons, jets, photons or invisible particles). Such signatures have been studied by the CMS collaboration with 3.7 \fbinv of \pp collisions data at \seighttev \cite{BH}. The $\ST={\sum \pt(objects)}$ variable is used to separate black hole candidate events from SM backgrounds, dominated by QCD multijet production, for several physical object multiplicities. As shown in Fig. \ref{fig:BH}.a, data are in agreement with the SM expectations and lower limits are set on black hole masses for several theoretical models (Fig. \ref{fig:BH}.b).

\subsection{Dark matter interpretations}

\begin{figure}
  \begin{center}
    \includegraphics[width=0.75\linewidth]{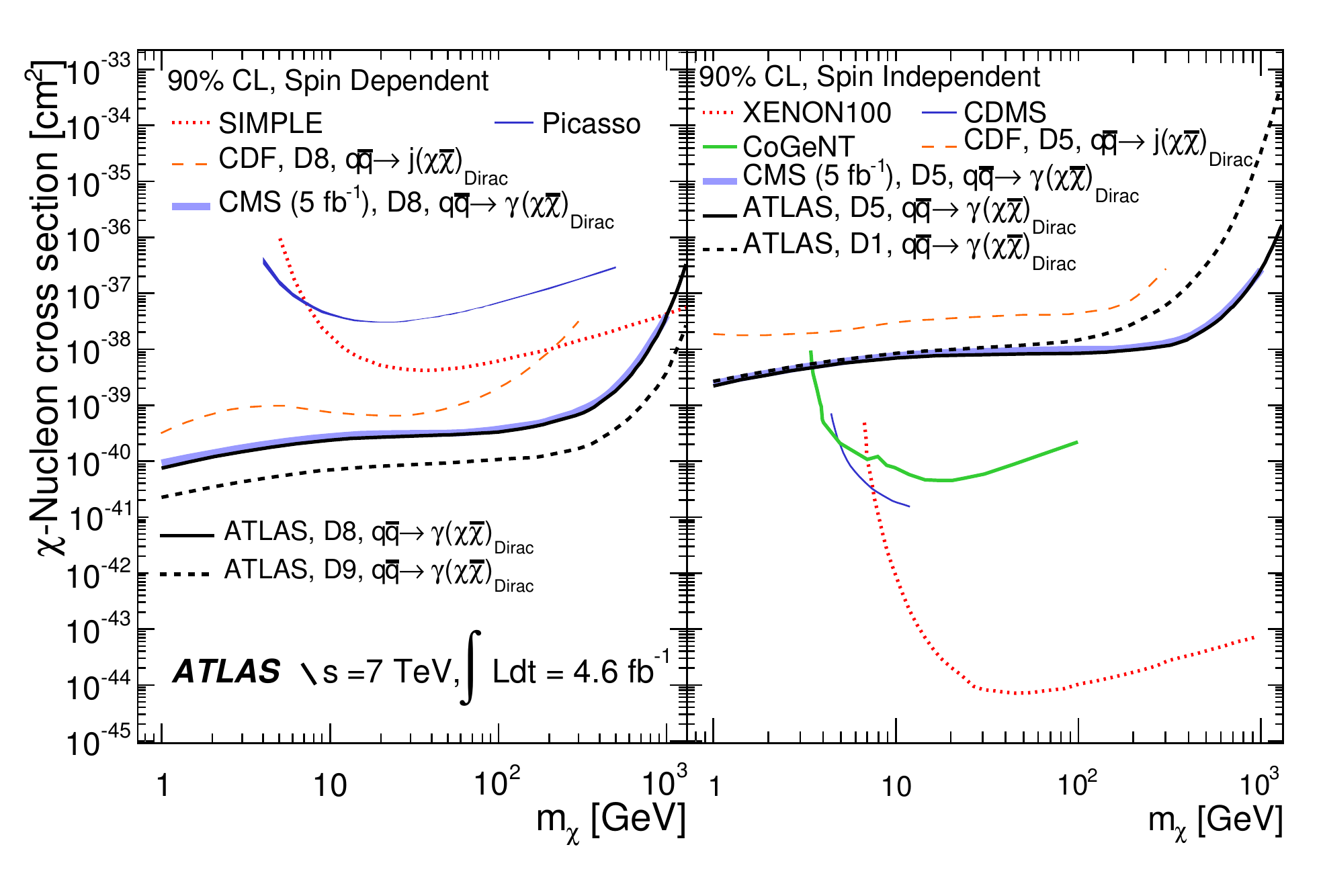}
    \caption{Limits on WIMPS-nucleon scattering cross section for several experiments, for a spin dependent interaction (right) and spin-independent interaction (left). }
    \label{fig:DM}
  \end{center}
\end{figure}

An interesting feature of the graviton production searches is their sensitivity to the production of dark matter particles $\chi$, also called WIMPs : instead of a graviton, a pair of WIMPs is produced in association with a jet or a photon. Therefore the monojet/monophoton results presented in section \ref{sec:EDGrav} can be interpreted as limits on the effective theory cut-off mass scale $M_{*}$ and as limit on the WIMP-nucleons scattering cross section \cite{MPhATLAS}. Several $qq\chi\chi$ interaction operators, spin-dependent or spin-independent have been considered. Fig. \ref{fig:DM} shows the upper limit on WIMP-nucleon inelastic cross section, for several experiments. One can notice that limits from collider experiments are the most stringent for low $\chi$ masses and are complementary to dedicated experiments.

\subsection{Contact interaction energy scale interpretation}

The study of dilepton signatures allow the experimentalists to probe also the presence of contact interaction. Since no excess in data have been observed (see section \ref{sec:EDGrav}), lower limits are set on the contact interaction energy scale $\Lambda$ by the ATLAS experiment \cite{DILATLAS} with a data sample of 5 \fbinv at \sseventev. Values of $\Lambda$ are excluded below 13.9 TeV at 95\% C.L. for constructive interferences while destructive interferences are excluded for $\Lambda < 10.2$ TeV at 95\% C.L..

\section{Searches for top resonances}




\subsection{ \ttbar resonances at high masses}

Several searches for high masses \ttbar resonances, produced by the decay of a high mass Z' boson or a Kaluza-Klein gluon, have been performed by the ATLAS and CMS collaboration with \pp collision data at \sseventev corresponding to 5 \fbinv in both semileptonic and full hadronic channels \cite{TTbHATLAS,TTbHCMS,TTbLATLAS,TTbLCMS}. No evidence of beyond Standard Model \ttbar resonances have been seen so far with \sseventev data. The most stringent limits have been set in the semileptonic channel (Table \ref{tab:ttbar}).

\begin{table}[!h]
  \begin{center}
    \begin{tabular}{|l||c|c|c|}
      \hline \hline
      model & narrow Z' ($\Gamma = 1\% M_{Z'}$) &  wide Z' ($\Gamma = 10\% M_{Z'}$)  & KK gluon \\
      \hline
      ATLAS & $0.5<\mbox{M}<1.7$ TeV & n/a & $0.7<\mbox{M}<1.9$ TeV \\
      CMS   & $0.5<\mbox{M}<1.5$ TeV & $0.5<\mbox{M}<2.0$ TeV & $1<\mbox{M}<1.82$ TeV \\
      \hline \hline
    \end{tabular}
    \label{tab:ttbar}
    \caption{More stringent limits on Z' and KK gluon masses from the search for \ttbar resonances set by ATLAS and CMS.}
  \end{center}
\end{table}

\subsection{top+jet resonances at high masses}

The ATLAS collaboration performed a search for top+jet resonances, produced by the decay of a high mass W' boson, in association with another top quark. The \pp collision data at \sseventev are used \cite{TJATLAS}. The search has been performed in the semileptonic channel. No excess of data have been observed in the reconstructed visible W' mass spectrum and limits were set on the W' cross section production, for different coupling benchmarks. For a unity left-handed coupling, W' are excluded for masses below 500 GeV at 95\% C.L..

\section{Searches for \ff generation signatures}

The discovery of the SM Higgs boson and related studies significantly disfavor the existence of a chiral \ff generation quark family. However, some other models are not ruled out and can still be probed by collider experiments.

\subsection{5/3e up quark signature}

The CMS experiment performed a search for 4$^{th}$ generation up quark with an electric charge 5/3e ($T_{5/3}$)  with a sample of \pp data at \seighttev corresponding to 20 \fbinv \cite{Ch4UCMS}. A $T_{5/3}$ decays into a W boson and a top quark. The presence of a pair of $T_{5/3}$ can then be probed into events containing two same sign leptons from the two W bosons produced by one $T_{5/3}$ and the corresponding top quark, and at least 5 jets with $\pt > 30$ GeV. The $\HT = \sum \pt(leptons,jets)$ variable is used to select signal events, by requiring $\HT > 900$ GeV. Fig. \ref{fig:up53} shows the \HT spectrum for data, SM backgrounds and a 600 GeV $T_{5/3}$ signal. No significant excess has been observed in the data and a lower limit has been set on the $T_{5/3}$ mass. At 95\% C.L., 5/3e charged 4$^{th}$ generation up quarks are excluded for masses below 700 GeV.


\subsection{Vector-like up quark signature}

\begin{figure}
  \begin{minipage}{0.49\linewidth}
    \begin{center}
      \includegraphics[width=0.99\linewidth]{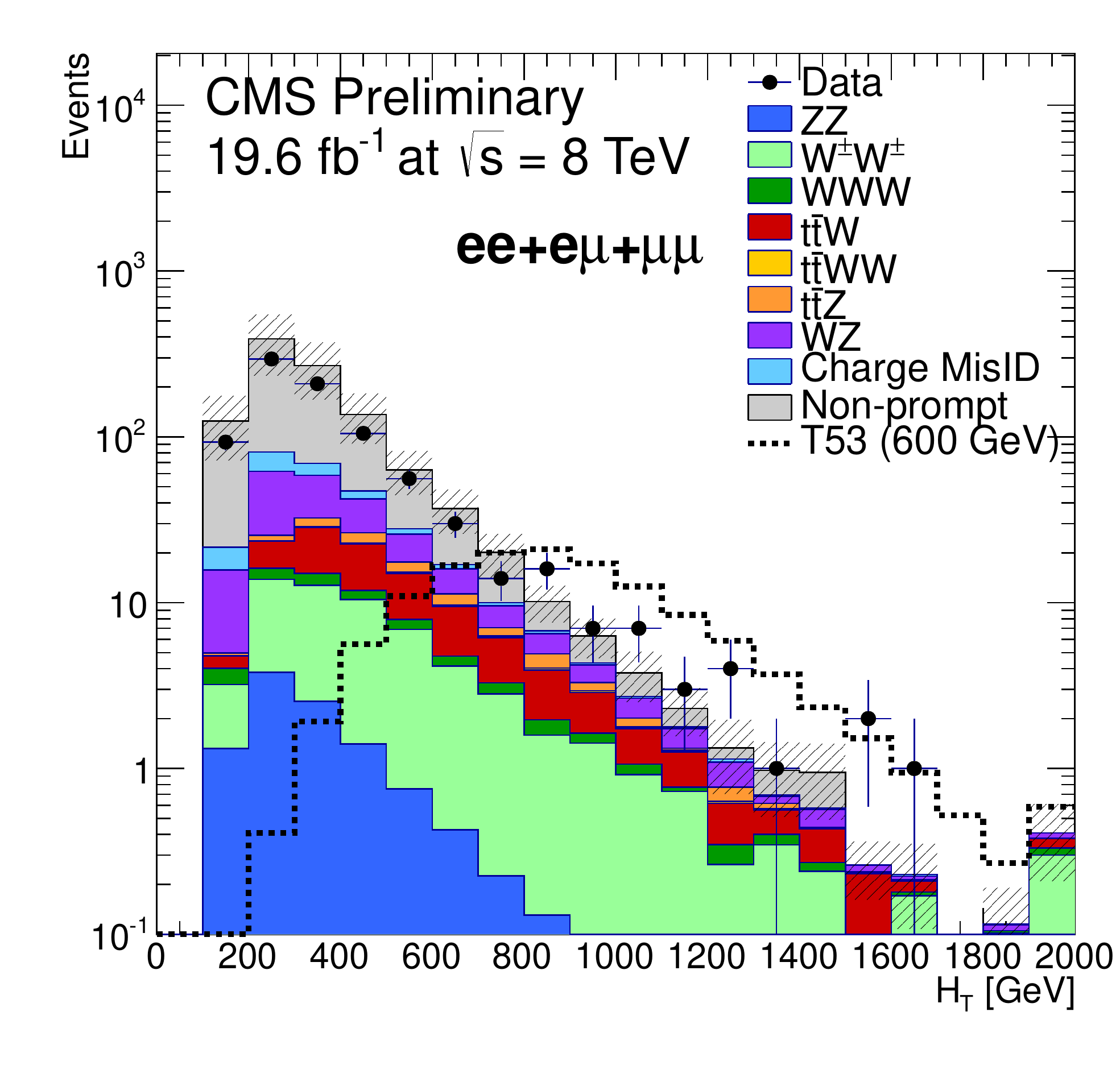}
      \caption{\HT distribution for data (black points), SM backgrounds (colored areas) and a 600 GeV $T_{5/3}$ signal.}
      \label{fig:up53}
    \end{center}
  \end{minipage}
  \hfill
  \begin{minipage}{0.49\linewidth}
    \begin{center}
      \includegraphics[width=0.99\linewidth]{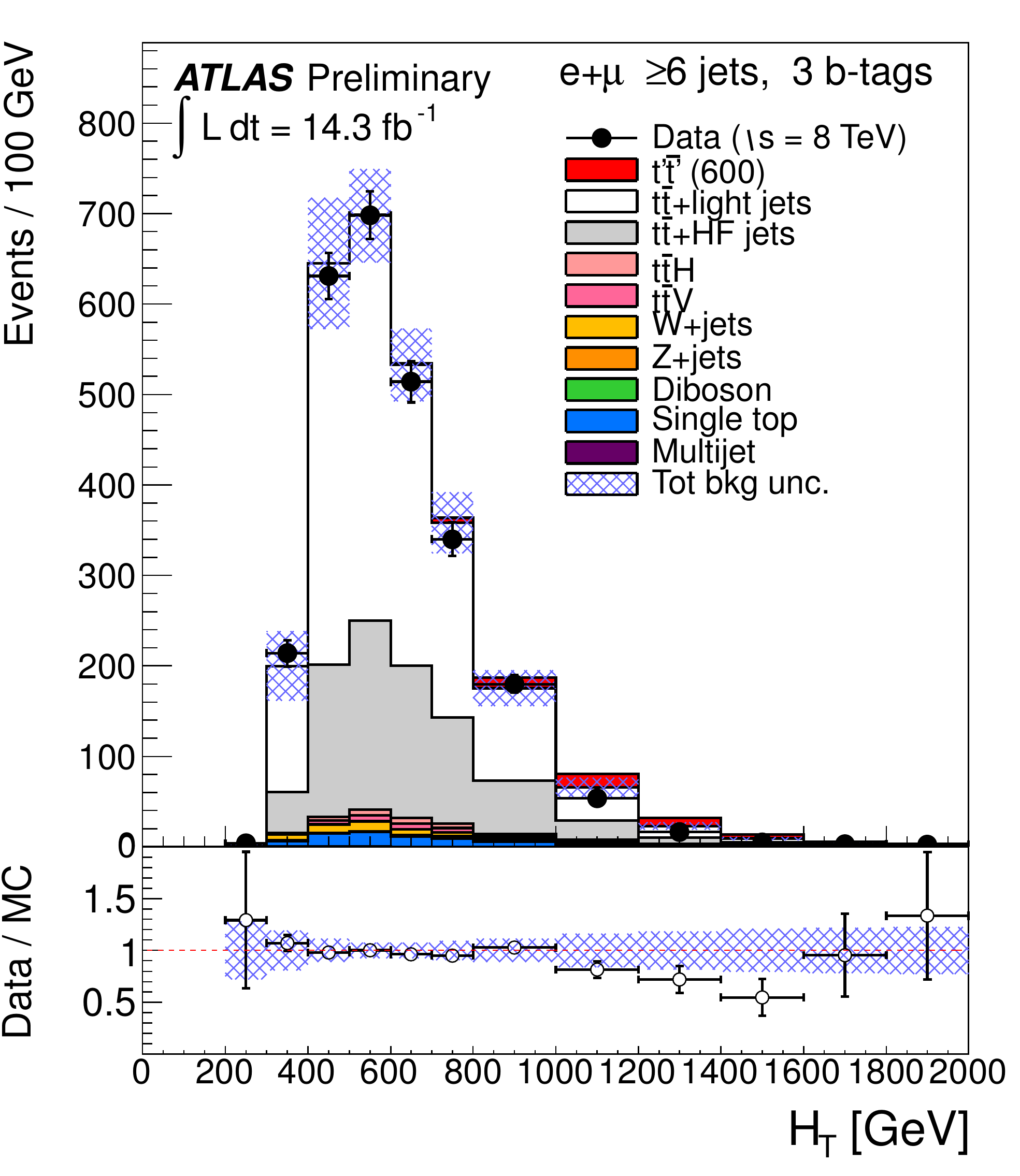}
      \caption{Scalar sum of transverse energies of lepton, \met and jets in the final state. Data are indicated by black points, signal by the red filled area and SM background expectations by white and colored areas. }
      \label{fig:upAtlas}
    \end{center}
  \end{minipage}
\end{figure}

A model of vector-like \ff generation up quark (VL-up quark) pair production has been probed by the ATLAS collaboration with \seighttev data with one VL-up quark decaying into a top quark and one SM Higgs boson $\mbox{H} \rightarrow \mbox{bb}$ while the second VL-up quark decays into a W (Z) boson and a b (t) quark \cite{VLUATLAS}. The semileptonic channel (1 charged lepton, large \met, at least 6 jets with at least two of them tagged as jets from b quarks) has been studied, by looking for a excess of data in the scalar sum of transverse energy \HT of the objects in the final state. As shown on Fig. \ref{fig:upAtlas}, data are in agreement with the SM expectations and the existence of weak isospin singlet (doublet) VL-up quarks is excluded at 95\% C.L. for masses below 640 (730) GeV.

\subsection{Excited down-quark signature}

The ATLAS collaboration also performed a search for excited down quarks (b$^{*}$)  \cite{DOQATLAS} with \pp data at \sseventev. The theoretical model used imply a chromomagnetic interaction between a b-quark and a gluon producing a b$^{*}$ decaying into a top quark and a W boson. Both semileptonic (1 electron or muon, \met, 3 jets) and dileptonic (2 electrons or muons, \met, one jet) channels have been studied. No excess with respect to SM expectations have been noticed and limits were set on the b$^{*}$ mass, for several couplings benchmarks. For a unity left-handed couplings, b$^{*}$ are excluded for masses below 870 GeV.


\section{Summary of leptoquark searches at $\sqrt{s}=7$ TeV}

An intriguing feature of the Standard Model is the symmetry between the number of quarks and lepton families. Several theories beyond the Standard Model (SU(5), grand unification theories, ...) predict the existence of new bosons, the leptoquarks (LQ), carrying a fractional electric charge, a color charge, and decaying into a lepton and a quark. Constraints from Flavor Changing Neutral Currents measurements significantly constraint a leptoquark to decay into a quark and a lepton of the same generation. The branching fraction $\beta$ of a LQ decay into a charged lepton and a quark is a free parameter of the model.

Searches for a pair production of scalar LQs of 1$^{st}$, 2$^{nd}$ and 3$^{rd}$ generation have been performed by both ATLAS and CMS experiments \cite{LQ1CMS,LQ2CMS,LQ1ATLAS,LQ2ATLAS,LQ3CMS,LQ3ATLAS} with \pp collision data at \sseventev corresponding to an integrated luminosity of 5 \fbinv . In all LQ decay channels, for all LQ generation, the data have been found to be compatible with SM expectations. The most stringent limits at 95\% C.L. set on leptoquark masses by the ATLAS and CMS collaborations are summarized in Table \ref{tab:LQ}.

\begin{table}[!h]
  \begin{center}
    \begin{tabular}{||l||c|c|c||}
      \hline \hline
  & $\beta=1$ & $\beta=0.5$ & $\beta=0$ \\
\hline
LQ1$\rightarrow eq(\nu_e q)$ & 830 & 640 & n/a \\
LQ2$\rightarrow \mu q(\nu_\mu q)$ & 840 & 650 & n/a \\
LQ3$\rightarrow \tau b(\nu_\tau b)$ & 538 & n/a & 450 \\
      \hline \hline
    \end{tabular}
    \caption{ Most stringent lower limits set on scalar leptoquark masses in GeV. $\beta$ is the branching fraction of a leptoquark decaying into a charged lepton and a quark.}
    \label{tab:LQ}
  \end{center}
\end{table}

\section{Conclusion}

Various models of physics beyond the Standard Model have been probed in various final states, and no evidence of new physics have been seen so far, with data collected by the ATLAS and CMS experiment at \sseventev or \seighttev. Limits were set on key observables of the different models. Further searches for new physics signatures are still ongoing with the full dataset recorded with \seighttev, corresponding to an total integrated luminosity of 20 \fbinv.

A complete list of results published by the ATLAS and CMS experiments can be found in the corresponding public twiki pages \cite{twikiATLAS,twikiCMS}.




\end{document}